\documentclass[preprint,showpacs,preprintnumbers,amsmath,amssymb]{revtex4}

\usepackage{graphicx}
\usepackage{dcolumn}
\usepackage{bm}
\usepackage{amssymb}

\begin{document}
\newcommand{\vecx}{\mbox{\boldmath $x$}}

\title{Generalized Fisher information matrix in nonextensive systems \\
with spatial correlation}

\author{Hideo Hasegawa}
\altaffiliation{hideohasegawa@goo.jp}
\affiliation{Department of Physics, Tokyo Gakugei University,  
Koganei, Tokyo 184-8501, Japan}%

\date{\today}

\begin{abstract}
By using the $q$-Gaussian distribution derived 
by the maximum entropy method for spatially-correlated $N$-unit 
nonextensive systems, we have calculated 
the generalized Fisher information matrix of $g_{\theta_n \theta_m}$ for  
$(\theta_1, \theta_2, \theta_3) = (\mu_q, \sigma_q^2$, $s$),
where $\mu_q$, $\sigma_q^2$ and $s$ denote the mean, variance and degree 
of spatial correlation, respectively, for a given entropic index $q$.
It has been shown from the Cram\'{e}r-Rao theorem that
(1) an accuracy of an unbiased estimate of $\mu_q$ is improved (degraded) 
by a negative (positive) correlation $s$,
(2) that of $\sigma_q^2$ is worsen with increasing $s$, and
(3) that of $s$ is much improved for $s \simeq -1/(N-1)$ or $s \simeq 1.0$
though it is worst at $s = (N-2)/2(N-1)$.
Our calculation provides a clear insight to the long-standing controversy
whether the spatial correlation is beneficial 
or detrimental to decoding in neuronal ensembles.
We discuss also a calculation of the $q$-Gaussian distribution, 
applying the superstatistics to the Langevin model subjected 
to spatially-correlated inputs. 
\end{abstract}

        

\maketitle
\newpage

\section{Introduction}

It is well known that the Fisher information plays an important 
role in statistical mechanics and information theory 
(for review see \cite{Frieden98}).
The Fisher information is a useful tool in evaluating an accuracy of
information decoding, providing the lower bound for estimation errors
of unbiased estimates in the Cram\'{e}r-Rao theorem \cite{Frieden98}.
The Fisher information expresses the metric tensor in the Riemannian space
spanned by the probability distribution functions (PDFs)
in the information geometry \cite{Amari00}.
Calculations of the Fisher information have been made for various systems such as 
neuronal ensembles \cite{Paradiso88}-\cite{Toyoizumi06}. 
Neurons in ensembles communicate information, emitting short voltage pulses 
called spikes, which propagate through axons and dendrites to
neurons in the next stage 
(for review see \cite{Rieke96}-\cite{Hasegawa07},
related references therein).
Main issues on the neuronal code are whether the information is encoded
in the rate of firings ({\it rate code}) or 
in the firing times ({\it temporal code}),
and whether the information is encoded in the activity 
of a single (or very few) neuron
or that of a large number of neurons ({\it population code}).
A recent success in brain-machine interface \cite{BMI} suggests 
that the population code for the firing rate
is employed in sensory and motor neurons, although it 
is still unclear what kinds of codes are adopted
in higher-level cortical neurons.

The theoretical study of the Fisher information has been
performed for a discussion on the accuracy of decoding
and the efficiency of information transmission 
\cite{Paradiso88}-\cite{Toyoizumi06}. 
Calculations of the Fisher information have been made
mainly for uncorrelated (independent) systems because of
a mathematical simplicity. 
It has been shown that in independent systems, 
the Fisher information increases proportionally to the ensemble size
\cite{Seung93,Abbott99,Sompolinsky01,Shamir01}.
However, the correlation 
among constituent elements is inevitable in real systems.
In neuronal ensembles, for example, statistical dependence 
among consisting neurons would be expected
because each neuron may receive the same external inputs
and because consisting neurons are generally interconnected 
\cite{Rieke96}-\cite{Hasegawa07}.
There has been a long-standing controversy how correlation affects
the efficiency of population coding.
Some researchers have shown that the correlation
enhances the effectiveness of neural population code 
\cite{Jenison00,Johnson04},
while some have claimed that the correlation hinders the population code
\cite{Gawne93,Zohary94,Abbott99,Panzeri99,Sompolinsky01,Shamir01}.
In particular, the Fisher information is shown to saturate 
to a finite value as the system size grows 
in the presence of a positive correlation
\cite{Abbott99,Sompolinsky01,Shamir01}.
This raises questions on the role of correlation
in information decoding.

In the last decade, much attention has been paid to the nonextensive
statistics since Tsallis proposed the so-called Tsallis entropy $S_q$.
For $N$-unit systems, $S_q$ is given by 
\cite{Tsallis88,Tsallis98,Tsallis01,Tsallis04}
\begin{eqnarray}
S_q &=& \frac{k_B}{q-1}\left( 1-\int p(\vecx)^q \:d \vecx \right),
\label{eq:A1}
\end{eqnarray}
where $q$ is the entropic index, $k_B$ the Boltzmann constant,
$\vecx = \{ x_i \}$ ($i=1$ to $N$), $d \vecx = \prod_{i=1}^N dx_i $,
and $p(\vecx)$ denotes the multivariate PDF.
In the limit of $q \rightarrow 1.0$, the Tsallis entropy 
given by Eq. (\ref{eq:A1}) reduces to
the Boltzmann-Gibbs-Shannon entropy,
\begin{eqnarray}
S_1 &=& - k_B \int p(\vecx) \: \ln p(\vecx)\:d \vecx.
\end{eqnarray}
The Tsallis entropy is non-additive because
for $p(A \cup B)=p(A)\:p(B)$, we obtain
\begin{eqnarray}
S_q(A \cup B) &=& S_q(A)+ S_q(B)-\frac{(q-1)}{k_B}S_q(A) S_q(B). 
\end{eqnarray}
The Tsallis entropy is super-extensive, extensive and sub-extensive 
for $q < 1$, $q=1$ and $q > 1$, respectively,
and $q-1$ expresses the degree of the nonextensivity. 
The PDF of $p(\vecx)$ in Eq. (\ref{eq:A1}) is obtained by using
the maximum entropy method (MEM) 
for the Tsallis entropy with some constraints. 
There are four possible MEMs at the moment:
original method \cite{Tsallis88},
un-normalized method \cite{Curado91}, normalized method \cite{Tsallis98}, 
and the optimal Lagrange multiplier (OLM) method \cite{Martinez00}.
The four methods are equivalent in the sense that distributions 
derived in them are easily transformed each other \cite{Ferri05}.
A comparison among the four MEMs is made in Ref. \cite{Tsallis04}.
The Tsallis entropy is a basis of the nonextensive statistics,
which has been successfully applied to
a wide class of systems with the long-range interaction
and/or non-equilibrium (quasi-equilibrium) states
\cite{Tsallis01,Tsallis04,Nonext}.

One of alternative approaches to the nonextensive statistics 
besides the MEM is the superstatistics \cite{Wilk00,Beck01,Beck05}
(for a recent review, see \cite{Beck07}).
In the superstatistics, it is assumed that 
{\it locally} the equilibrium state of a given system
is described by the Boltzmann-Gibbs statistics and
its global properties may be expressed by a superposition 
over the fluctuating intensive parameter ({\it i.e.,} the inverse temperature) 
\cite{Wilk00}-\cite{Beck07}.
The superstatistics has been adopted in many kinds 
of subjects such as hydrodynamic turbulence \cite{Beck03,Reynolds03,Beck07b}, 
cosmic ray \cite{Beck04} and solar flares \cite{Baies06}.

The generalized Fisher information (GFI) in the nonextensive statistics
is defined by \cite{Abe03}-\cite{Hasegawa08a}
\begin{eqnarray}
g_{\theta_n \theta_m} 
&=& q\: E\left[ \left( \frac{\partial \ln p(\vecx)}{\partial \theta_n} \right)
\left( \frac{\partial \ln p(\vecx)}{\partial \theta_m} \right) \right],
\label{eq:A2}
\end{eqnarray}
where $E[\cdot]$ stands for the expectation value over the PDF of
$p(\vecx)$ [$=p(\vecx \vert \theta )$],
and $\theta$ parameters specifying the PDF.
Equation (\ref{eq:A2}) is derived from 
the generalized Kullback-Leibler divergence which is in conformity with 
the Tsallis entropy \cite{Abe03}-\cite{Hasegawa08a}.
In the limit of $q \rightarrow 1.0$, the GFI
given by Eq. (\ref{eq:A2}) reduces to the conventional one.
In a previous paper \cite{Hasegawa08b}, 
we discussed the effect of the spatial correlation on
the Tsallis entropy and the GFI,
calculating $S_q$ and $g_{\theta \theta}$ for $\theta=\mu_q$,
where $\mu_q$ stands for mean value [Eq. (\ref{eq:B4})]. 
It is the purpose of the present paper to extend the calculation 
to the GFI matrix of $g_{\theta_n \theta_m}$ for 
$(\theta_1, \theta_2, \theta_3)=(\mu_q, \sigma_q^2$, $s$),
where $\sigma_q^2$ and $s$ express variance and degree 
of the spatial correlation, respectively [Eqs. (\ref{eq:B5}) 
and (\ref{eq:B6})].
We will investigate the dependence of the GFI on $s$, $N$ and $q$,
by using the PDF derived by the OLM-MEM \cite{Martinez00}.
Such detailed calculations of the GFI matrix 
have not been reported even for the extensive system ($q=1.0$),
as far as the author is aware of. The calculated GFI 
is expected to provide us with a clear insight to the controversy 
on a role of the spatial correlation discussed above.
Quite recently, we have pointed out the possibility that
input information to neuronal ensembles may be carried not only 
by mean but also by variance and/or correlation
in firing rate within the population code hypothesis
\cite{Hasegawa08c,Hasegawa08d}.
The inverse of the calculated GFI matrix expresses
an accuracy of decoding when input information is carried
by such population codes.

The paper is organized as follows. In Sec. II, we obtain
the PDF by the OLM-MEM for spatially-correlated nonextensive systems.
In Sec. III, the maximum likelihood estimator for the inference
of the parameters is discussed.
In Sec. IV, analytic expressions for elements of the GFI matrix 
are presented with some model calculations.
In Sec. V, the PDF for the Langevin model with spatially-correlated inputs
is calculated within the superstatistics \cite{Wilk00,Beck01}, which is
compared to that derived by the MEM in Sec. II.
Section VI is devoted to conclusion with 
the relevance of our calculation to decoding in neuronal population code
\cite{Hasegawa08c,Hasegawa08d}.

\section{Maximum entropy method}
\subsection{Probability distribution function}

We consider spatially-correlated $N$-unit nonextensive systems,
for which the Tsallis entropy is given by 
Eq. (\ref{eq:A1}) \cite{Tsallis88,Tsallis98}.
We derive the PDF, $p(\vecx)$, by using the OLM-MEM \cite{Martinez00}
for the Tsallis entropy, imposing the constraints given by \cite{Hasegawa08b}
\begin{eqnarray}
1 &=& \int p(\vecx)\:d \vecx, 
\label{eq:B3}
\\
\mu_q &=& \frac{1}{N}\sum_i E_q\left[ x_i \right], 
\label{eq:B4}\\
\sigma_q^2 &=& \frac{1}{N} \sum_i
E_q\left[(x_i-\mu_q)^2 \right], 
\label{eq:B5} \\
s \:\sigma_q^2 &=& \frac{1}{N(N-1)}\sum_i \sum_{j (\neq i)}
E_q\left[(x_i-\mu_q)(x_j-\mu_q) \right],
\label{eq:B6}
\end{eqnarray}
where 
$E_q[\cdot]$ denotes an expectation value averaged over the escort distribution function
of $P_q(\vecx)$,
\begin{eqnarray}
P_q(\vecx) &=& \frac{p(\vecx)^q}{\int p(\vecx)^q \:d \vecx}.
\label{eq:B7}
\end{eqnarray}
The OLM-MEM with the constraints given by 
Eqs. (\ref{eq:B3})-(\ref{eq:B6}) leads to the PDF
given by (for details, see Appendix B 
of Ref. \cite{Hasegawa08b})
\begin{eqnarray}
p(\vecx) &=& \frac{1}{Z_q}
\exp_q\left[- \left( \frac{1}{2 \nu_q \sigma_q^2 }\right)
\sum_{i=1}^{N} \sum_{j=1}^{N} C_{ij} 
(x_i-\mu_q)(x_j-\mu_q) \right],
\label{eq:C1}
\end{eqnarray}
with
\renewcommand{\arraystretch}{1.5}
\begin{eqnarray}
Z_q = \left\{ \begin{array}{ll}
r_s \left(\frac{2\nu_q \sigma_q^2}{q-1} \right)^{N/2}
\;\; \prod_{i=1}^N 
\:B\left(\frac{1}{2}, \frac{1}{q-1}-\frac{i}{2} \right) 
\quad & \mbox{for $1 < q <3$}, \label{eq:C5}\\
r_s (2 \pi \sigma_q^2)^{N/2}
\quad & \mbox{for $q=1$}, \label{eq:C6} \\
r_s \left(\frac{2 \nu_q \sigma_q^2}{1-q}\right)^{N/2}
\;\; \prod_{i=1}^N 
\:B\left(\frac{1}{2}, \frac{1}{1-q}+\frac{(i+1)}{2} \right)
\quad & \mbox{for $ q <1$}, \label{eq:C7} \\
\end{array} \right.
\end{eqnarray}
\begin{eqnarray}
C_{ij} &=& c_0 \;\delta_{ij}+ c_1 \:(1-\delta_{ij}), 
\label{eq:C2} \\
c_0 &=& \frac{[1+(N-2)s]}{(1-s)[1+(N-1)s]}, \label{eq:C3} \\
c_1 &=& - \:\frac{s}{(1-s)[1+(N-1)s]}, \label{eq:C4} \\
r_s &=& \: \{(1-s)^{N-1}[1+(N-1)s] \}^{1/2},
\label{eq:C8} \\ 
\nu_q &=& \frac{[(N+2)-Nq]}{2},
\label{eq:C9}
\end{eqnarray}
where $B(p,q)$ denotes the beta function and
$\exp_q(x)$ the $q$-exponential function defined by
\begin{eqnarray}
\exp_q(x) &=& 
[1+(1-q)x]_{+}^{1/(1-q)},
\label{eq:C10}
\end{eqnarray}
with $[x]_{+} ={\rm max}(x,0)$. We hereafter assume that 
the entropic index $q$ takes a value,
\begin{eqnarray}
0 < q < 1+\frac{2}{N} \leq 3,
\label{eq:C12}
\end{eqnarray}
because $p(\vecx)$ given by Eq. (\ref{eq:C1}) has the probability properties
with $\nu_q > 0$ for $q < 1+2/N $ 
and because the Tsallis entropy is stable for $q > 0$ 
\cite{Abe02}.

In the limit of $q=1.0$, the PDF given by Eq. (\ref{eq:C1}) 
becomes the multivariate Gaussian distribution given by
\begin{eqnarray}
p(\vecx) &=& \frac{1}{Z_1}
\exp\left[-\left(\frac{1}{2 \sigma_1^2}\right) 
\sum_{ij} C_{ij}(x_i-\mu_1)(x_j-\mu_1) \right].
\label{eq:C11}
\end{eqnarray}

\section{Maximum likelihood estimator}

The logarithmic likelihood estimator for $M$ sets of data 
of $\vecx_m=\{x_{im}\}$ ($i=1$ to $N$, $m=1$ to $M$) is given by

\begin{eqnarray}
\ln L(\theta) &=& \sum_{m=1}^M \ln p(\vecx_k \vert \theta)
= - \left( \frac{1}{q-1} \right) \sum_{m=1}^M \ln U(\vecx_m) 
-M \ln Z_q,
\label{eq:D1}
\end{eqnarray}
with
\begin{eqnarray}
U(\vecx_m) &=& 1+\frac{(q-1)}{2 \nu_q \sigma_q^2} 
\sum_{ij} C_{ij} (x_{im}-\mu_q) (x_{jm}-\mu_q).
\label{eq:D2}
\end{eqnarray}
Variational conditions for parameters of
$\theta=\mu_q,\:\sigma_q^2$ and $s$ lead to
\begin{eqnarray}
\frac{\partial \ln L}{\partial \mu_q}
&=& \frac{1}{\nu_q \sigma_q^2} \sum_{m=1}^M 
\sum_{ij} \frac{C_{ij} (x_{im}-\mu_q)}{U(\vecx_m)} =0, 
\label{eq:D3} \\
\frac{\partial \ln L}{\partial \sigma_q^2}
&=& \frac{1}{2\nu_q \sigma_q^4} \sum_{m=1}^M \sum_{ij}  
\frac{C_{ij}(x_{im}-\mu_q)(x_{jm}-\mu_q)}{U(\vecx_m)}
-\frac{MN}{2  \sigma_q^2} =0, 
\label{eq:D4} \\
\frac{\partial \ln L}{\partial s}
&=& - \frac{1}{2\nu_q \sigma_q^2} \sum_{m=1}^M \sum_{ij} 
\frac{ (dC_{ij}/ds) (x_{im}-\mu_q)(x_{jm}-\mu_q)}{U(\vecx_m)} \nonumber \\
&+& \frac{MN(N-1)}{2(1-s)[1+(N-1)s]}=0, 
\label{eq:D5}
\end{eqnarray}
After some calculations using Eqs. (\ref{eq:C2})-(\ref{eq:C4}),
(\ref{eq:D3})-(\ref{eq:D5}), we obtain
\begin{eqnarray}
\mu_q &=& \frac{\sum_m \sum_i x_{im} U(\vecx_m)^{-1} }
{N \sum_m 1/U(\vecx_m)^{-1} }, 
\label{eq:D6} \\
\sigma_q^2 &=& \frac{1}{\nu_q M N } \sum_m \sum_i 
\frac{(x_{im}-\mu_q)^2}{U(\vecx_m)},
\label{eq:D7} \\
s \:\sigma_q^2 &=& \frac{1}{\nu_q M N(N-1)} 
\sum_m \sum_{i}\sum_{j (i\neq j)} 
\frac{(x_{im}-\mu_q) (x_{jm}-\mu_q)}{U(\vecx_m)},
\label{eq:D8}
\end{eqnarray}
from which $\mu_q$, $\sigma_q^2$ and $s$ are 
self-consistently determined.

In the case of $q=1.0$, Eqs. (\ref{eq:D6})-(\ref{eq:D8}) become
\begin{eqnarray}
\mu_1 &=& \frac{1}{MN} \sum_m \sum_i x_{im},
\label{eq:D9} \\
\sigma_1^2 &=& \frac{1}{M N } \sum_m \sum_i (x_{im}-\mu_1)^2,
\label{eq:D10} \\
s \:\sigma_1^2 &=& \frac{1}{M N(N-1)} 
\sum_m \sum_{i}\sum_{j (i\neq j)} 
(x_{im}-\mu_1)(x_{jm}-\mu_1).
\label{eq:D11}
\end{eqnarray}

\section{Generalized Fisher information}

We have calculated elements of the GFI matrix 
given by Eq. (\ref{eq:A2}) with a basis of 
$(\theta_1, \theta_2,\theta_3)=(\mu_q, \sigma_q^2, s)$, 
as given by (for details, see the Appendix)
\[\sf {G}=\left(
\begin{array}{ccc}
{\displaystyle \frac{N}{\sigma_q^2 [1+(N-1)s]}} & 0 & 0\\
0 &  {\displaystyle \frac{N \nu_q} {2 \sigma_q^4} }  
& {\displaystyle - \frac{N(N-1)\nu_q \:s}{2\sigma_q^2(1-s)[1+(N-1)s]} } \\
0 & {\displaystyle - \frac{N(N-1)\nu_q \:s}{2\sigma_q^2(1-s)[1+(N-1)s]} }
& {\displaystyle \frac{N(N-1)[1+(N-1)\nu_q s^2]}
{2(1-s)^2[1+(N-1)s]^2} }
\end{array}
\right).\]
\begin{eqnarray}
\label{eq:E1}
\end{eqnarray}
The positive definiteness of $g_{\theta \theta}$ 
in Eq. (\ref{eq:E1}) imposes 
the condition on conceivable values of $s$ and $q$ given by
\begin{eqnarray}
-\frac{1}{(N-1)}&\equiv&s_L < s \leq 1,
\label{eq:E2} \\
q &\leq& 1+\frac{2}{N},
\label{eq:E3}
\end{eqnarray}
The physical origin of Eq. (\ref{eq:E2}) is expressed by
(see Appendix C in Ref. \cite{Hasegawa08b})
\begin{eqnarray}
0 &\leq & E_q[(X-\mu_q)^2] 
\leq \frac{1}{N} \sum_{i} \:E_q[(x_i-\mu_q)^2]=\sigma_q^2, 
\end{eqnarray}
which signifies that the global fluctuation in $X$ ($=N^{-1} \sum_i \:x_i$) 
is smaller than the average of local fluctuations in $\{ x_i \}$.
The condition given by Eq. (\ref{eq:E3}) is satisfied by $q$ in Eq. (\ref{eq:C12}).

In the limit of $q=1.0$ where $\nu_q=1.0$, 
Eq. (\ref{eq:E1}) reduces to
\[\sf {G}=\left(
\begin{array}{ccc}
{\displaystyle \frac{N}{\sigma_1^2 [1+(N-1)s]} } & 0 & 0\\
0 & {\displaystyle \frac{N}{2 \sigma_1^4} }
& {\displaystyle -\frac{N(N-1)s}{2\sigma_1^2(1-s)[1+(N-1)s]} } \\
0 & {\displaystyle -\frac{N(N-1)s}{2\sigma_1^2(1-s)[1+(N-1)s]} }
& {\displaystyle \frac{N(N-1)[1+(N-1)s^2]}{2(1-s)^2[1+(N-1)s]^2} }
\end{array}
\right),\]
\begin{eqnarray}
\end{eqnarray}
which is in agreement with the result obtained directly from the multivariate
Gaussian distribution given by Eq. (\ref{eq:C11}).

In the limit of $s=0.0$ ({\it i.e.,} no correlation), the GFI 
matrix given by Eq. (\ref{eq:E1}) becomes 
\begin{eqnarray}
\sf {G}&=&\left(
\begin{array}{ccc}
{\displaystyle \frac{N}{\sigma_q^2}} & 0 & 0\\
0 &  {\displaystyle \frac{N \nu_q} {2 \sigma_q^4} }  
& 0  \\
0 & 0
& {\displaystyle \frac{N(N-1)}
{2} }
\end{array}
\right),
\end{eqnarray}
whose elements of $g_{\mu_q \mu_q}$ and $g_{\sigma_q^2 \sigma_q^2}$
agree with those obtained previously in Ref. \cite{Hasegawa08a}. 

The Cram\'{e}r-Rao theorem implies that 
the lower bound of an unbiased estimate
of the parameters is expressed by
the inverse of the GFI matrix, which is given by
\[\sf {G^{-1}}=\left(
\begin{array}{ccc}
{\displaystyle \frac{\sigma_q^2 [1+(N-1)s]}{N} } & 0 & 0\\
0 &  {\displaystyle \frac{2\sigma_q^4[1+(N-1)\nu_q s^2]}{N \nu_q} }  
& {\displaystyle \frac{2 \sigma_q^2 \:s(1-s)[1+(N-1)s]}{N} } \\
0 & {\displaystyle \frac{2 \sigma_q^2 \:s(1-s)[1+(N-1)s]}{N}}
& {\displaystyle \frac{2 (1-s)^2 [1+(N-1)s]^2}{N(N-1)} }
\end{array}
\right). \]
\begin{eqnarray}
\label{eq:E4}
\end{eqnarray}

\begin{figure}
\begin{center}
\includegraphics[keepaspectratio=true,width=100mm]{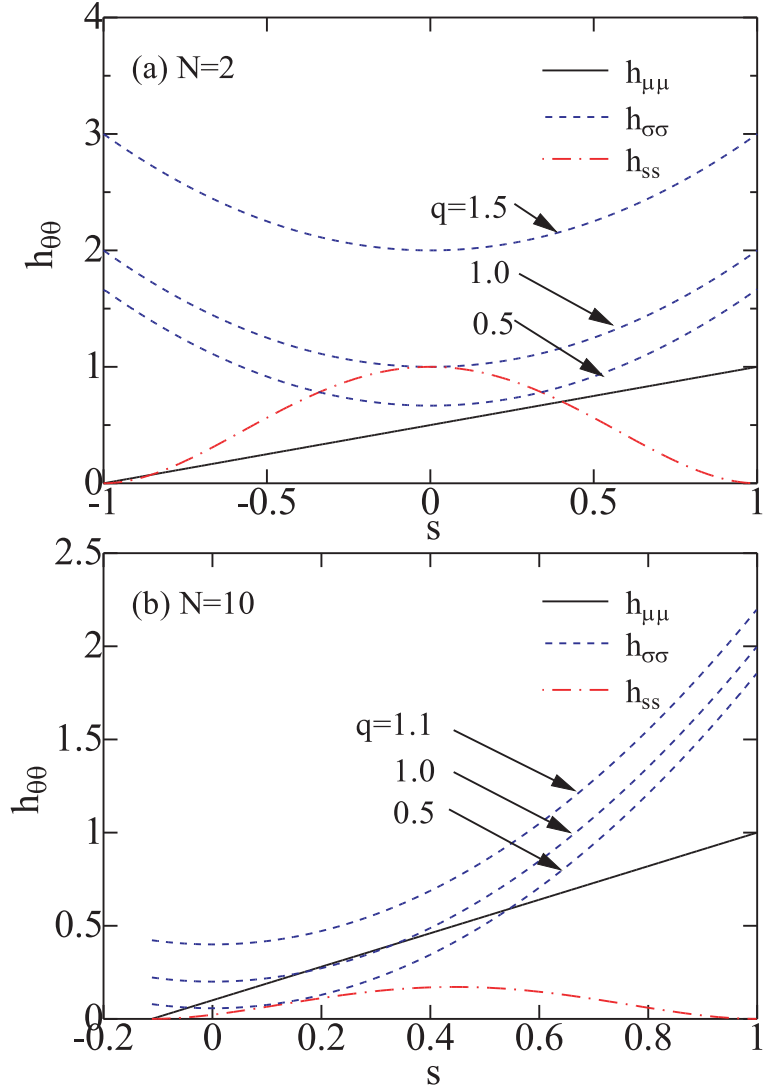}
\end{center}
\caption{
(Color online)
The $s$ dependence of inverses of the GFI, 
$h_{\mu_q \mu_q}$ (solid curves), $h_{\sigma_q^2 \sigma_q^2}$ (dashed curves)
and $h_{ss}$ (chain curves),
with (a) $N=2$ and (b) $N=10$ for various $q$ ($\mu_q=0.0$ and $\sigma_q^2=1.0$).
}
\label{fig1}
\end{figure}

\begin{figure}
\begin{center}
\includegraphics[keepaspectratio=true,width=100mm]{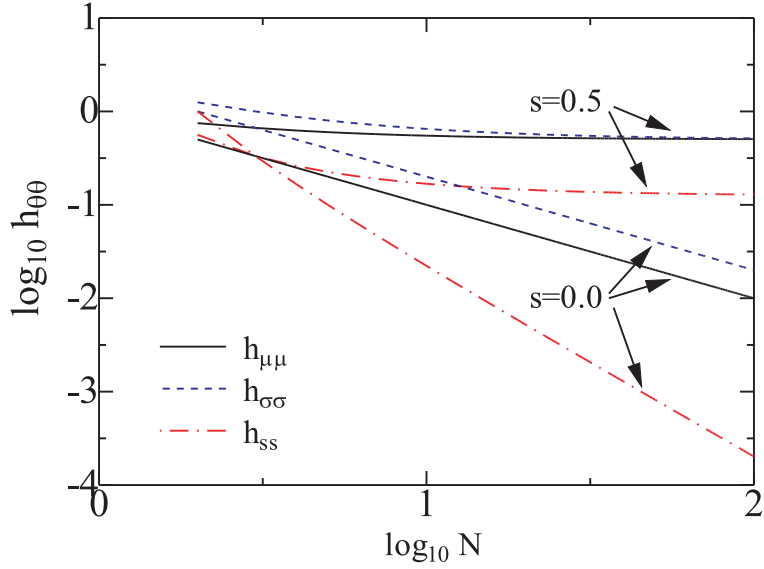}
\end{center}
\caption{
(Color online)
The $N$ dependence of inverses of the GFI,
$h_{\mu_q \mu_q}$ (solid curves), $h_{\sigma_q^2 \sigma_q^2}$ (dashed curves)
and $h_{ss}$ (chain curves),
for $s=0.0$ and $s=0.5$ ($q=1.0$, $\mu_q=0.0$ and $\sigma_q^2=1.0$).
}
\label{fig2}
\end{figure}

\begin{figure}
\begin{center}
\includegraphics[keepaspectratio=true,width=100mm]{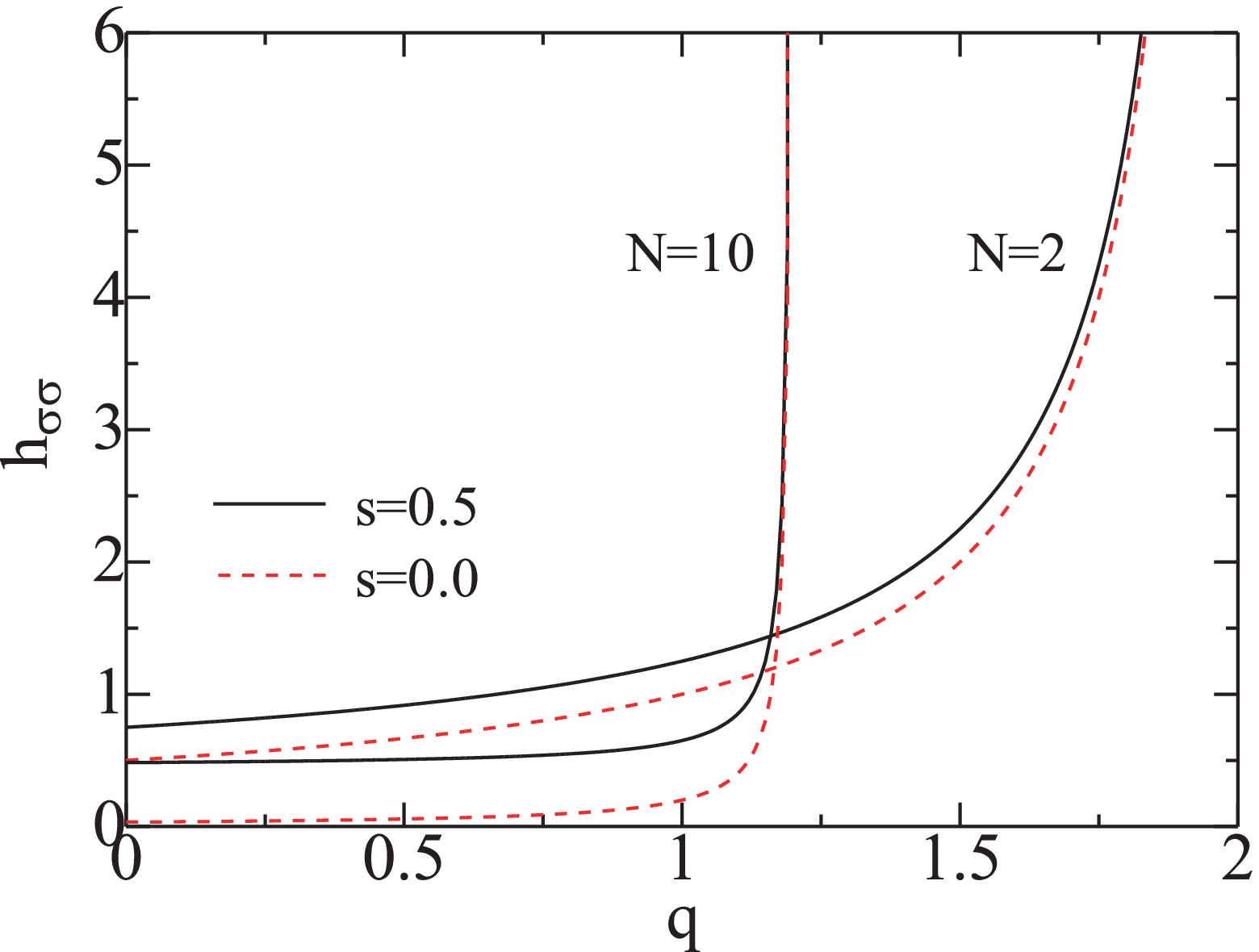}
\end{center}
\caption{
(Color online)
The $q$ dependence of inverses of the GFI,
$h_{\sigma_q^2 \sigma_q^2}$, 
for $s=0.0$ (dashed curves) and $s=0.5$ (solid curves)
with $N=2$ and $N=10$ ($\mu_q=0.0$ and $\sigma_q^2=1.0$).
}
\label{fig3}
\end{figure}

Equations (\ref{eq:E1}) and (\ref{eq:E4}) are 
the main result of our study.
In what follows, we examine the $s$, $N$ and $q$ dependence of
the inversed GFI matrix of 
$h_{\theta \theta} \equiv ({\sf G^{-1}})_{\theta \theta}$
with some model calculations which are presented 
in Figs. \ref{fig1}-\ref{fig3}.  

\vspace{0.5cm}
\noindent
{\bf The $s$ dependence}

Equation (\ref{eq:E4}) shows (1) $h_{\mu_q \mu_q}=0.0$ at $s=s_L$, 
(2) $h_{\sigma_q^2 \sigma_q^2}$ has a minimum at $s=0.0$,
and (3) $h_{ss}$ vanishes at $s=s_L$ and $s=1.0$.
The maximum of $h_{ss}$ locates at $s=(N-2)/2(N-1) \equiv s_M $,
which becomes $s_M = 0.5$ for a large $N$.
Figure \ref{fig1}(a) shows the $s$ dependence of the inverse of
the GFI for $N=2$ which is expressed by
\begin{eqnarray}
\sf {G^{-1}}=\left(
\begin{array}{ccc}
{\displaystyle \frac{\sigma_q^2 (1+s)}{2} } & 0 & 0\\
0 &  {\displaystyle \frac{\sigma_q^4(1+\nu_q s^2)}{\nu_q} }  
& {\displaystyle \sigma_q^2 \:s(1-s^2) } \\
0 & {\displaystyle \sigma_q^2 \:s(1-s^2)}
& {\displaystyle (1-s^2)^2 }
\end{array}
\right). 
\label{eq:E5}
\end{eqnarray}
With increasing $s$ from $s=s_L=-1.0$, $h_{\mu_q \mu_q}$ is linearly
increased.
$h_{ss}$ and $h_{\sigma_q^2 \sigma_q^2}$ are symmetric with respect to $s=0.0$
where $h_{ss}$ ($h_{\sigma_q^2 \sigma_q^2}$) has a maximum (minimum).
Figure 1(b) shows a similar plot for $N=10$ for which $s_L=-0.11$.
With increasing $s$ from $s=-0.11$, $h_{\mu_q \mu_q}$ is linearly
increased.
$h_{ss}$ has a maximum at $s=s_M=0.44$ and
vanishes at $s=-0.11$ and $s=1.0$.

\vspace{0.5cm}
\noindent
{\bf The $N$ dependence}

We note in Eq. (\ref{eq:E4}) that for $s=0$, the GFI is proportional to $N$.
For a finite positive $s$, however, they show the saturation when $N$ is increased: 
for $N \rightarrow \infty$, we obtain $h_{\mu_q \mu_q}=\sigma_q^2 s$, 
$h_{\sigma_q^2 \sigma_q^2}=2 \sigma_q^4 s^2$ and $h_{ss}=2s^2 (1-s)^2$.
For a negative $s$, inverse matrix elements tend to vanish
as $N$ approaches $(1+\vert s \vert)/\vert s \vert$.
The calculated $N$ dependence of $h_{\theta \theta}$ is plotted
in Fig. \ref{fig2}, where inversed  matrix elements 
for $s=0.5$ saturate at $N \gtrsim 10$, although those for $s=0.0$ 
is proportional to $N^{-1}$.  

\vspace{0.5cm}
\noindent
{\bf The $q$ dependence}

Equation (\ref{eq:E4}) shows that
$h_{\mu_q \mu_q}$ and $h_{ss}$ are independent of $q$, while
$h_{\sigma_q^2 \sigma_q^2}$ is increased with increasing $q$ from $q=0$.
This increase is due to a factor of $\nu_q$ in Eq. (\ref{eq:C9}), 
which is decreased with increasing $q$ and which diverges at $1+2/N$: 
note that $\nu_q=N/2+1$, 1.0 and 0.0 for $q=0.0$,
$q=1.0$ and $q=1+2/N$, respectively.
The calculated $q$ dependence of $h_{\sigma_q^2 \sigma_q^2}$ is plotted
in Fig. \ref{fig3}, where it diverges at $q=2.0$ ($q=1.2$)
for $N=2$ ($N=10$).

\section{Discussion}
We have discussed the GFI 
for the $q$-Gaussian distribution derived by the MEM 
\cite{Tsallis88,Tsallis98,Martinez00}.
It is possible to derive the $q$-Gaussian distribution
by using the Langevin model within the superstatistics 
\cite{Wilk00,Beck01}. 
We consider an ensemble consisting of $N$ elements in a given system.
The dynamics of a variable $x_i$ ($i=1$ to $N$) is assumed to be described by
the Langevin model given by
\begin{eqnarray}
\frac{dx_{i}}{dt} &=& - \lambda x_i+I_i(t). 
\label{eq:F1}
\end{eqnarray}
Here $\lambda$ denotes the relaxation rate and
input signals $I_i(t)$ have variability defined by
\begin{eqnarray}
I_i(t) &=& I(t) + \delta I_i(t), 
\label{eq:F2}
\end{eqnarray}
with
\begin{eqnarray}
\langle \delta I_i(t) \rangle &=& 0, 
\label{eq:F3} \\
\langle \delta I_i(t) \delta I_j(t') \rangle 
&=& 2 D [\delta_{ij}  + s_I (1-\delta_{ij}) ]
\delta(t-t'),
\label{eq:F4}
\end{eqnarray}
where the bracket $\langle \cdot \rangle$ signifies the ensemble average, 
and $2 D$ and $s_I$ denote the variance and degree of 
the spatial correlation, respectively. 
The variability in Eq. (\ref{eq:F4}) arises from noise and/or
heterogeneity in consisting elements.
The origin of the spatial correlation may be common 
external inputs and/or couplings among elements.


The Fokker-Planck equation for the PDF of $\pi(\vecx,t)$ for
$\vecx=\{x_i \}$ is given by
\begin{eqnarray}
\frac{\partial \pi(\vecx,t)}{\partial t}
&=& \sum_i \frac{\partial }{\partial x_i}[\lambda x_i- I(t)]\:\pi(\vecx,t)
+ D \sum_i \sum_j Q_{ij} 
\frac{\partial^2}{\partial x_i \partial x_j}\:\pi(\vecx,t), 
\label{eq:F5}
\end{eqnarray}
with the covariance matrix {\sf Q} whose elements are given by
\begin{eqnarray}
Q_{ij} &=& \delta_{ij}+s_I (1-\delta_{ij}).
\end{eqnarray}

The solution of the FPE (\ref{eq:F5}) is given by
\begin{eqnarray}
\pi(\vecx,t) = \left( \:\frac{1}{r_s\: [2\:\pi \sigma^2]^{N/2} } \right)
\exp\left(-\frac{1}{2 \sigma^2} 
\sum_i \sum_j C_{ij}\:(x_i-\mu)(x_j-\mu) \right),
\label{eq:F6}
\end{eqnarray}
where $\mu$, $\sigma^2$ and $s$ obey equations of motion given
(argument $t$ being suppressed) 
\begin{eqnarray}
\frac{d \mu}{d t} &=& -\lambda \mu + I,
\label{eq:F7} \\
\frac{d \sigma^2}{d t}
&=& -2 \lambda \sigma^2 + 2 D,  
\label{eq:F8} \\
\frac{d s}{d t}
&=& -\left( \frac{2D}{\sigma^2} \right) (s-s_I),
\label{eq:F9} 
\end{eqnarray}
$C_{ij}$ and $r_s$ being defined by Eqs. (\ref{eq:C2}) and (\ref{eq:C8}),
respectively.
We note in Eqs. (\ref{eq:F7})-(\ref{eq:F9}) that $\mu(t)$ is decoupled from
$\sigma^2(t)$ and $s(t)$, and that $\sigma^2(t)$ is independent of
$s(t)$ although $s(t)$ depends on $\sigma^2(t)$.
In the stationary state, we obtain
\begin{eqnarray}
\mu &=& I/\lambda, \;\;\;
\sigma^2 = \frac{D}{\lambda}, \;\;\;
s=s_I. 
\end{eqnarray}


After the concept in the superstatistics \cite{Wilk00,Beck01,Beck05,Beck07},
we assume that a model parameter of $\beta$ ($= 1/\sigma^2=\lambda/D$)
fluctuates, and that its distribution is expressed by 
the $\chi^2$-distribution with rank $n$, 
\begin{eqnarray}
f(\beta) &=& \frac{1}{\Gamma(n/2)}\left(\frac{n}{2\beta_0} \right)^{n/2}
\beta^{n/2-1} e^{-n\beta/2 \beta_0}
\hspace{1cm}\mbox{($n=1,2,\cdot \cdot \cdot$)},
\label{eq:G1}
\end{eqnarray}
where $\Gamma(x)$ is the gamma function.
Average and variance of $\beta$ are given by
$ \langle \beta \rangle_{\beta}=\beta_0 $
and $(\langle \beta^2 \rangle_{\beta}-\beta_0^2)/\beta_0^2=2/n$, respectively.
Taking the average of $\pi(\vecx)$ over $f(\beta)$,
we obtain the stationary distribution given by 
\begin{eqnarray}
p(\vecx)&=& \int_0^{\infty} \pi(\vecx) \:f(\beta)\:d\beta,\\  
& = & \frac{1}{Z_q}
\exp_{q} \left(- \frac{1}{2 \gamma_q} 
\sum_i\sum_j C_{ij} (x_i -\mu)(x_j-\mu) \right),
\label{eq:G2}
\end{eqnarray}
with
\begin{eqnarray}
Z_q &=& \left\{ \begin{array}{ll}
r_s\:\left( \frac{2 \gamma_q}{q-1}\right)^{N/2}
\;\prod_{i=1}^N B\left(\frac{1}{2}, \frac{1}{q-1}-\frac{i}{2} \right) 
\quad & \mbox{for $q > 1$},
\label{eq:G3} \\
r_s \:(2 \pi \gamma_q)^{N/2}
\quad & \mbox{for $q=1$}, 
\label{eq:G4} 
\end{array} \right. \\
q &=& 1+ \frac{2}{(N+n)}, 
\label{eq:G5} \\
\gamma_q &=& \frac{n}{\beta_0 \:(N+n)}
= \frac{(N+2)-N q}{2 \beta_0},
\end{eqnarray}
where $r_s$ is given by Eq. (\ref{eq:C8}).
In the limit of $n \rightarrow \infty$ ($q \rightarrow 1.0$) where 
$f(\beta) \rightarrow \delta(\beta-\beta_0)$, 
the PDF reduces to the multivariate Gaussian distribution 
given by
\begin{eqnarray}
p(\vecx)  
&= & \frac{1}{Z_1}\exp \left(- \frac{\beta_0}{2}\: 
\sum_i\sum_j C_{ij}(x_i -\mu)(x_j-\mu) \right),
\label{eq:G6}
\end{eqnarray}
which agrees with Eq. (\ref{eq:F6}) 
for $\beta_0=\lambda/D=1/\sigma^2$.

We note that the PDF given by Eq. (\ref{eq:G2}) is equivalent to that
given by Eq. (\ref{eq:C1}) derived by the MEM 
when we read $\mu=\mu_q$ and $\gamma_q=\nu_q \sigma_q^2$,
besides the fact that the former is defined for $1 \leq q \leq [1+2/(N+n)] <2$
[Eq. (\ref{eq:G5})]
while the latter for $0 < q < (1+2/N) < 3$ [Eq. (\ref{eq:C12})].

In the limit of $s=0$ ({\it i.e.,} no spatial correlation),
Eq. (\ref{eq:G2}) reduces to
\begin{eqnarray}
p(\vecx)  
& \propto  & \exp_{q} \left(- \frac{1}{2 \gamma_q }
\sum_i (x_i -\mu)^2 \right),\\
&\propto & p(x_1) \otimes_q p(x_2) \otimes_q \cdot\cdot \otimes_q \: p(x_N),
\label{eq:G7}
\end{eqnarray}
with
\begin{eqnarray}
p(x_i) & \propto  & \exp_q \left(- \frac{1}{2 \gamma_q }
(x_i -\mu)^2 \right),
\label{eq:G8}
\end{eqnarray}
where the $q$-product is defined by \cite{Borges04}
\begin{eqnarray}
x \otimes_q y &=& [x^{1-q}+y^{1-q}-1]^{1/(1-q)}.
\label{eq:G9}
\end{eqnarray}
Note that in deriving Eq. (\ref{eq:G7}), the normalization factors
of $p(x_i)$ are not taken into account.


\section{Conclusion}
We have calculated the GFI
matrix in spatially-correlated nonextensive systems.
From the Cram\'{e}r-Rao theorem, the calculated GFI 
implies the followings: 
(i) an accuracy of an estimate of $\mu_q$ is improved (degraded) 
by a negative (positive) correlation,
(ii) that of $\sigma_q^2$ is worsen with increasing $s$,
(iii) that of $s$ is much improved for $s \simeq -1/(N-1)$ 
and $s \simeq 1.0$ while it is worst at $s=s_M = (N-2)/2(N-1)$,
(iv) those of all parameters are improved with increasing $N$, and
(v) that of $\sigma_q^2$ is worsen with increasing $q$ at 
$ q > 1$ and its estimation is impossible for $q \geq 1+2/N$, 
while those of $\mu_q$ and $s$ are independent of $q$.

The points (i) and (iv) are consistent with
previous results for extensive systems ($q=1.0$)
\cite{Abbott99,Panzeri99,Sompolinsky01}.
The point (iii) shows that if input information is carried by synchrony
within the population code hypothesis 
\cite{Hasegawa08c,Hasegawa08d},
its decoding accuracy may be improved either by small or large correlation,
independently of $q$ [the point (v)].
Our calculation concerns the long-standing controversy on a role of
the synchrony in neuronal ensembles \cite{Gawne93}-\cite{Johnson04}. 

\begin{acknowledgments}
This work is partly supported by
a Grant-in-Aid for Scientific Research from the Japanese 
Ministry of Education, Culture, Sports, Science and Technology.  
\end{acknowledgments}

\vspace{0.5cm}
\appendix*

\section{Calculations of the generalized
Fisher information matrix}

First we express PDFs of $p(\vecx)$ in Eq. (\ref{eq:C1}) 
in a compact form given by
\begin{eqnarray}
p(\vecx)  
&=& \frac{U^{-b} }{Z_q},
\label{eq:X1}
\end{eqnarray}
with
\begin{eqnarray}
U &=& 1+ a^2 \sum_i \sum_j C_{ij}(x_i-\mu_q)(x_j-\mu_q),
\label{eq:X2} 
\end{eqnarray}
\begin{eqnarray}
Z_q = \left\{ \begin{array}{ll}
\frac{r_s}{a^N}
\;\prod_{i=1}^N B\left(\frac{1}{2}, b-\frac{i}{2} \right) 
\quad & \mbox{for $1 < q < 3$}, \\
r_s (2 \pi \sigma_q^2)^{N/2} 
\quad & \mbox{for $q = 1$}, \\
\frac{r_s}{a^N}
\;\prod_{i=1}^N B\left(\frac{1}{2}, -b+\frac{(i+1)}{2} \right) 
\quad & \mbox{for $q < 1$}, 
\end{array} \right.
\end{eqnarray}
\begin{eqnarray}
a &=& 
\left( \frac{\vert q-1 \vert}{2 \nu_q \sigma_q^2} \right)^{1/2}, \\
b &=& \frac{1}{q-1}. 
\end{eqnarray}
By using the unitary transformation, Eq. (\ref{eq:X2}) is
transformed to
\begin{eqnarray}
U &=& 1+a^2 \sum_i \lambda_i \:y_i^2, 
\end{eqnarray}
where $\lambda_i$ and $y_i$ express eigen-values 
and eigen-vactors, respectively.
We obtain $\lambda_i$ given by
\begin{eqnarray}
\lambda_i &=& \frac{1}{[1+(N-1)s]} 
\hspace{2cm}\mbox{for $i=1$},\\
&=& \frac{1}{(1-s)}
\hspace{2cm}\mbox{for $1< i \leq N$}. 
\end{eqnarray}
Explicit expressions for $y_i$ are not necessary
for our discussion, except for $y_1$ given by
\begin{eqnarray}
y_1 &=& \frac{1}{\sqrt{N}} \sum_i (x_i-\mu_q).
\end{eqnarray}
Taking the derivatives of $\ln p(\vecx)$ with respect to
parameters of $\mu_q$, $\sigma_q^2$ and $s$, and
performing tedious calculations with Eq. (\ref{eq:A2}),
we may obtain the GFI matrix elements given by
Eq. (\ref{eq:E1}). 
In deriving them, we have employed the 
following expectation values:
\begin{eqnarray}
E\left[ \frac{1}{U} \right] 
&=& \frac{(b-N/2)}{b}, \\
E\left[ \frac{y_i^2}{U}  \right]
&=& \frac{1}{2 a^2 b \lambda_i}, \\
E\left[ \frac{y_i^2}{U^2} \right]
&=& \frac{(b-N/2)}{2a^2 b(b+1) \lambda_i}, \\
E\left[ \frac{y_i^4}{U^2} \right] 
&=& \frac{3 }{4a^4 b(b+1) \lambda_i^2}, \\
E\left[ \frac{y_i^2 y_j^2}{U^2} \right] 
&=&  \frac{1}{4a^4 b(b+1) \lambda_i \lambda_j}
\hspace{1cm}\mbox{for $i \neq j$},
\end{eqnarray}
where $E[\cdot]$ denotes the average over $p(\vecx)$.




\end{document}